\documentclass[prl,amsmath,amssymb,twocolumn]{revtex4-1}
\usepackage{graphicx}
\usepackage{subfigure}
\usepackage{adjustbox}
\usepackage{bm}
\usepackage{color}
\usepackage{braket}
\usepackage{standalone}
\usepackage{multirow}
\usepackage{tikz}
\usepackage{mathrsfs}
\usepackage{dsfont}

\usepackage[colorlinks,bookmarks=true,citecolor=blue,linkcolor=red,urlcolor=blue]{hyperref}

\newcommand{\bea}{\begin{eqnarray}}
\newcommand{\eea}{\end{eqnarray}}

\begin{document}

\title{Gate-Tunable Fractional Chern Insulators in Twisted Double Bilayer Graphene}

\author{Zhao Liu$^1$}
\email{zhaol@zju.edu.cn}
\author{Ahmed Abouelkomsan$^2$}
\email{ahmed.abouelkomsan@fysik.su.se}
\author{Emil J. Bergholtz$^3$}\email{emil.bergholtz@fysik.su.se}
\affiliation{$^1$Zhejiang Institute of Modern Physics, Zhejiang University, Hangzhou 310027, China\\
$^2$Department of Physics, Stockholm University, AlbaNova University Center, 106 91 Stockholm, Sweden}
\date{\today}

\begin{abstract} 
	We predict twisted double bilayer graphene to be a versatile platform for the realization of fractional Chern insulators readily targeted by tuning the gate potential and the twist angle.
	Remarkably, these topologically ordered states of matter, including spin singlet Halperin states and spin polarized states in Chern number $\mathcal C=1$ and $\mathcal{C}= 2$ bands, occur at high temperatures and without the need for an external magnetic field.
\end{abstract}

\maketitle

{\it Introduction.} Following the remarkable discovery of superconductivity and correlated states in magic angle twisted bilayer graphene \cite{twistedbilayermodel0,twistedbilayermodel1,cao2018unconventional,cao2018correlated,yankowitz2019tuning,Lu2019}, understanding the phase diagram resulting from electron-electron interactions in different Moir\'e heterostructures have attracted the interest of many.
As a natural progression to the study of the acclaimed twisted bilayer graphene, recent investigations have geared towards twisted double bilayer graphene (TDBG) systems \cite{PhysRevB.99.075127,ashvinTDBG,jeilTDBG,alexnanoletters,koshinoTDBG,choiTDBG}, formed when two bilayers of graphene---instead of monolayers---are rotated with respect to each other. Experiments on TDBG have established it as a promising platform for interaction-driven states. In particular it has been reported \cite{correlatedTDBG1,correlatedTDBG2,pablotdbgnature,correlatedTDBG3} that spin polarized correlated insulator states arise in the narrow flat conduction band.

An important class of states that could arise due to the strong electron-electron interactions in an isolated flat band is fractional Chern insulators (FCIs) \cite{Emilreview,Sidreview}. These are lattice analogs of the conventional fractional quantum Hall (FQH) effect in two-dimensional electron gases, with the advantage that they occur at zero magnetic field and with potentially significantly larger gaps. 
There has been a growing interest in investigating different kinds of FCIs in Moir\'e systems~\cite{OurFCI,CecileFCI,AshvinFCI,yvesexcitonic,intiexcitonic,soluyanovFQH}. Most saliently, recent theoretical studies pointed to the existence of valley polarized FCIs at fractional filling of the topological $|\mathcal{C}|= 1$ valence or conduction band of twisted bilayer graphene aligned with boron nitride~\cite{OurFCI,CecileFCI,AshvinFCI}, as well as their competing gapless phases induced by strong particle-hole symmetry breaking terms which make the fine-tuning of experimental parameters necessary~\cite{OurFCI}. 

\begin{figure}
	\centerline{\includegraphics[width=1\linewidth]{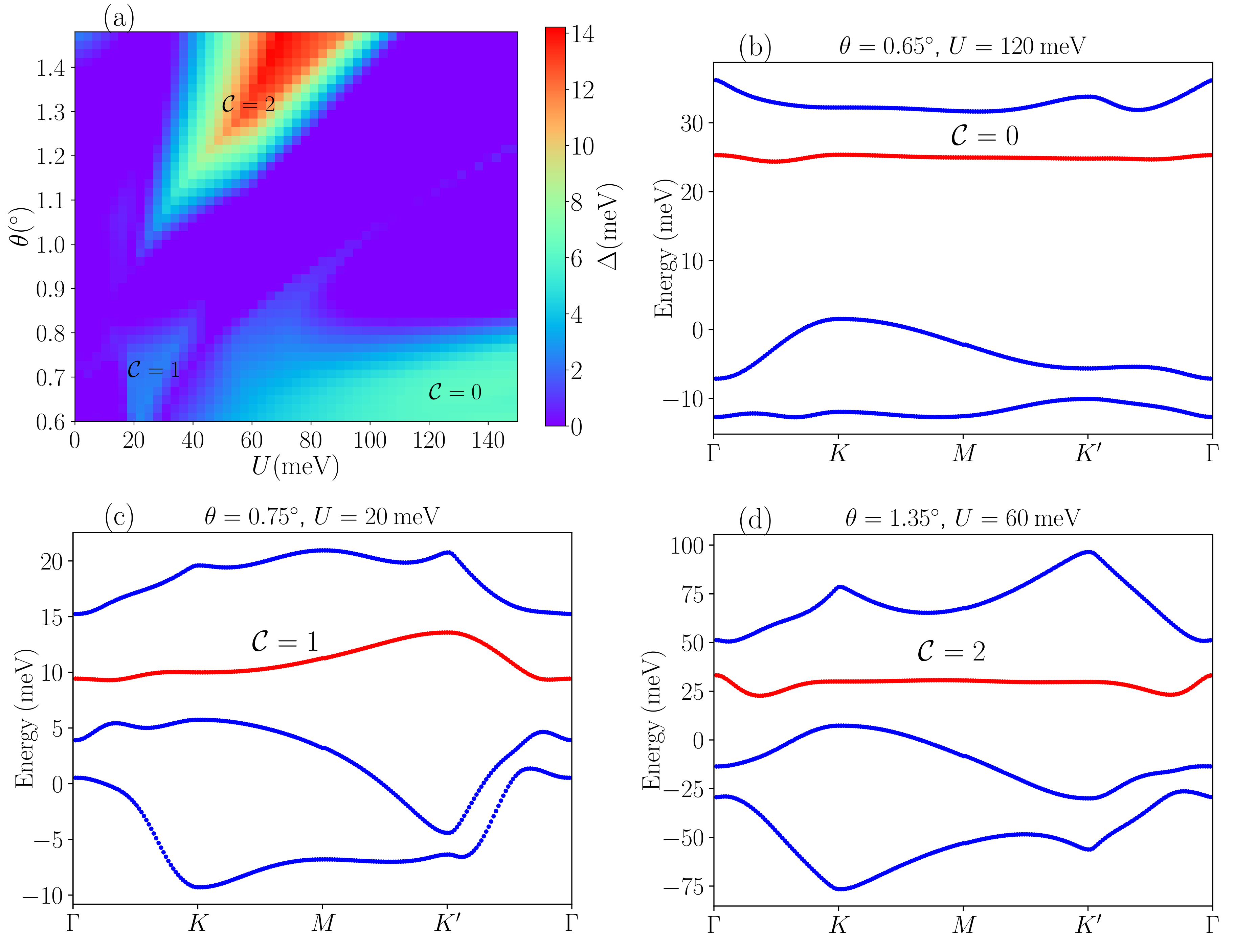}}
	\caption{(a) The indirect gap of the first conduction band as a function of the vertical voltage bias $U$ and the twist angle $\theta$. The Chern number $\mathcal{C}$ of the first conduction band is given in the three main regions where the band is isolated. (b)-(d) Non-interacting band structures near charge neutrality for selective values of $\theta$ and $U$ along $\Gamma=(0,0)$, $K = K_M(\sqrt{3}/2,1/2) $, $M = K_M(\sqrt{3}/4,3/4)$, and $K' = K_M (0,1)$ with $K_M = \frac{4\pi}{3 a_M}$. The first conduction band is shown in red with its Chern number $\mathcal{C}$. Plots are generated with $(t_0,t_1,t_3,t_4,\delta,w_1,w_0)=(2610,361,283,138,15,100,0.7w_1){\rm meV}$ \cite{supple}.} 
	\label{TDBG_gap}
\end{figure} 
\begin{figure*}[htb]
	\centerline{\includegraphics[width=\linewidth]{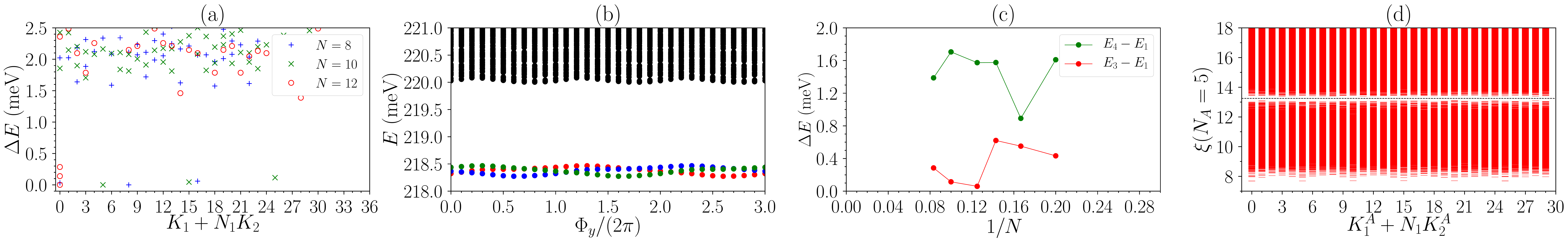}}
	\caption{Evidence of $\nu=1/3$ FCIs in the $\mathcal{C}=1$ region with $\theta=0.75^\circ$, $U=20{\rm meV}$. (a) The low-lying energy spectrum for $N=8, N_1\times N_2=4\times6$, $N=10,N_1\times N_2=5\times6$ and $N=12,N_1\times N_2=6\times6$. (b) The spectral flow for $N=10,N_1\times N_2=5\times 6$, where $\Phi_y$ is the magnetic flux insertion in the ${\bf a}_2$-direction. (c) The finite-size scaling of the energy gap (green) and the ground-state splitting (red) for $N=4,5,6,8,10,12$ without flux insertions. We define the energy gap and the ground-state splitting as $E_4-E_1$ and $E_3-E_1$, respectively, where $E_i$ is the $i$th energy level in ascending order. (d) The particle entanglement spectrum for $N=10,N_1\times N_2=5\times 6,N_A=5$, with $23256$ levels below the entanglement gap (the dashed line).}
	\label{C_1_nu_1_3}
\end{figure*}

While very encouraging, twisted bilayer graphene aligned with boron nitride only provides flat bands with $|\mathcal{C}|=1$, where FCIs can be directly mapped to conventional FQH states due to the topological equivalence between $|\mathcal{C}|=1$ bands and Landau levels ~\cite{qiwannier,gaugefix,zhaowannier}. By contrast, a single band with higher Chern number cannot be simply mapped to multiple decoupled Landau levels~\cite{highCWu}, leading to novel FCIs in $|\mathcal{C}|>1$ flat bands that do not have conventional FQH states as continuum counterparts~\cite{highCWang,highCLiu,max,highCYang,highCSterdyniak,highCMoller,highCWu}. These $|\mathcal{C}|>1$ FCIs have a ``color-entangled'' nature which is absent for conventional multicomponent FQH states~\cite{highCWu}, thus providing a richer category of FQH physics. However, they were not reported yet in Moir\'e systems without an external magnetic field. 

Motivated by these findings and open questions, here we explore the premises for FCIs in TDBG. TDBG serves as a natural platform for probing FCIs for a number of reasons. First, its layer configuration breaks the $C_2$ symmetry by default, hence removing the band touching at the Dirac points without needing to add a substrate layer. This results in separated bands around charge neutrality that could be individually studied. Second, 
TDBG bandwidth is highly controllable by tuning the twist angle $\theta$ and more significantly via the application of an electric field $U$, providing an extra degree of tunability in experimental setups that are absent in twisted bilayer graphene systems in the regime of twist angles $\theta = 0.6^{\circ}-1.4^{\circ} $ that we consider. This is to be contrasted with twisted bilayer graphene in the regime of very tiny twist angles $\theta \ll 1^{\circ}$ where electric fields have a qualitatively different impact on the properties \cite{tinytwist1,tinytwist2,tinytwist3,tinytwist4,tinytwist5,tinytwist6}. In TDBG the band topology is also controlled by varying $U$ and $\theta$. In particular, the first conduction band, which we focus on, has Chern number values ranging between $\mathcal{C}=0$ and $\mathcal{C}=3$ \cite{PhysRevB.99.075127,ashvinTDBG,jeilTDBG}, allowing us access not only to Landau levels alike bands with $|\mathcal{C}|= 1$ but also to $|\mathcal{C}| > 1$ bands that are topologically distinct from continuum Landau levels. The appearance of an easily accessed $\mathcal{C}=2$ flat band opens the possibility of realizing novel $\mathcal{C}=2$ FCIs~\cite{highCWang,highCLiu,max,highCYang,highCSterdyniak,highCMoller,highCWu} in realistic materials without an external magnetic field. 

In this Letter, we provide compelling evidence for the existence of FCIs in TDBG through a detailed microscopic study of projected Coloumb interactions onto the first conduction band for both $\mathcal{C}=1$ and $\mathcal{C}=2$ regimes. We find a variety of ferromagnetic and spin singlet FCIs at different filling factors that are highly tunable i.e, could be accessed by moving around in the $U-\theta$ space.

\begin{figure*}[htb]
	\centerline{\includegraphics[width=\linewidth]{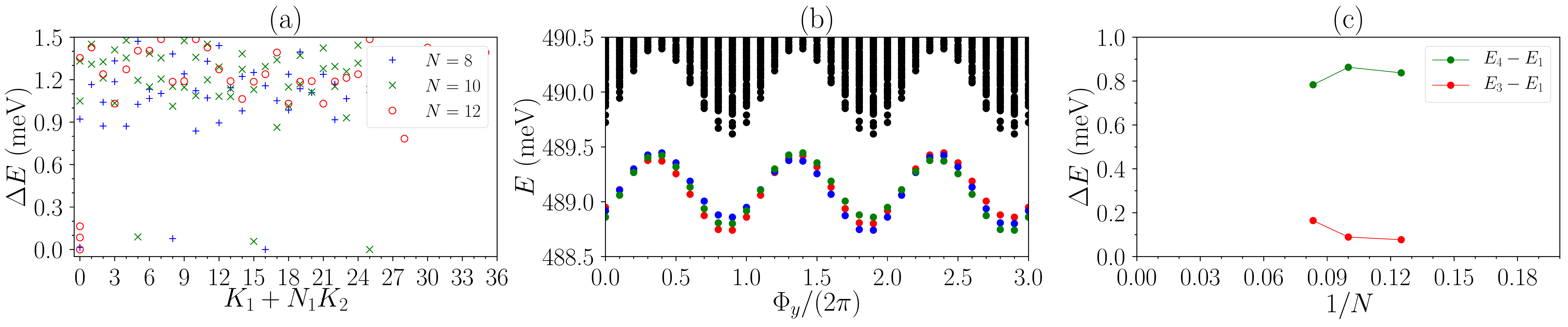}}
	\caption{Evidence of $\nu=1/3$ FCIs in the $\mathcal{C}=2$ region with $\theta=1.35^\circ$, $U=60{\rm meV}$. (a) The low-lying energy spectrum for $N=8, N_1\times N_2=4\times6$, $N=10,N_1\times N_2=5\times6$ and $N=12,N_1\times N_2=6\times6$. (b) The spectral flow for $N=10,N_1\times N_2=5\times 6$, where $\Phi_y$ is the magnetic flux insertion in the ${\bf a}_2$-direction. (c) The finite-size scaling of the energy gap (green) and the ground-state splitting (red) for $N=8,10,12$ without flux insertions. We define the energy gap and the ground-state splitting as $E_4-E_1$ and $E_3-E_1$, respectively, where $E_i$ is the $i$th energy level in ascending order.}
	\label{C_2_nu_1_3}
\end{figure*}

{\em Setup.} We consider electrons interacting via the screened Coulomb potential in the TDBG Moir\'e superlattice. We choose the Yukawa potential $V({\bf q})=\frac{e^2}{4\pi\epsilon_r\epsilon_0 S}\frac{2\pi}{\sqrt{|{\bf q}|^2+\kappa^2}}$ to describe the screening of the Coulomb interaction, where $e$ is the electron charge, $\epsilon_0$ is the dielectric constant of vacuum, $\epsilon_r\approx 4$ is the relative dielectric constant of the material~\cite{jeilTDBG}, $S$ is the area of the Moir\'e superlattice, and $\kappa$ measures the screening strength which we set as $\kappa=1/a_M$ with $a_M$ the lattice constant of TDBG~\cite{supple}. We emphasize that our results are not sensitive to the details of the screening mechanism~\cite{supple}. Unless otherwise stated, we assume both polarized spin and valley degrees of freedom for electrons. When the electrons are doped above the charge neutrality point, the first conduction band is partially filled. If the first conduction band is isolated from other bands below and above, it is fair to project the total Hamiltonian to this active band, leading to
\begin{eqnarray}
\label{onesector}
H^{\rm proj} = \sum_{{\bf k}}E({\bf k}) c_{\bf k}^\dagger c_{\bf k}+\sum_{\{\mathbf{k}_i\}} V_{\mathbf{k}_1\mathbf{k}_2\mathbf{k}_3\mathbf{k}_4} c^\dagger_{\mathbf{k}_1} c^\dagger_{\mathbf{k}_2} c_{\mathbf{k}_3} c_{\mathbf{k}_4},
\end{eqnarray} 
where $E({\bf k})$ is the dispersion of the first conduction band, $c^\dagger_{\mathbf{k}}$ ($c_{\mathbf{k}}$) creates (annihilates) an electron with momentum $\mathbf{k}$ in the first conduction band (per spin and valley), and all ${\bf k}_i$'s are in the Moir\'e Brillouin zone (MBZ). The matrix element $V_{\mathbf{k}_1\mathbf{k}_2\mathbf{k}_3\mathbf{k}_4}$ can be derived based on the effective model of TDBG as detailed in the Supplemental Material~\cite{supple}. To obtain a realistic model of TDBG, we include trigonal warping and particle-hole asymmetry terms, and choose $w_1=100{\rm meV}$ and $w_0/w_1=0.7$ throughout this work, where $w_0$ and $w_1$ are the AA and AB tunneling strengths between two sheets of bilayer graphene, respectively~\cite{supple}. Despite that the band dispersion is often neglected in studies of FCIs in other settings in order to emphasize the interaction effects, here we keep this term for considering a more realistic situation. In what follows, we impose periodic boundary conditions on finite samples and use extensive exact diagonalization to study the low-energy properties of the Hamiltonian (\ref{onesector}) at various filling factors $\nu = N / (N_1 N_2)$, where $N$ is the number of electrons in the first conduction band and $N_1$ and $N_2$ are the number of unit cells in the two basic directions ${\bf a}_1$ and ${\bf a}_2$ of the TDBG Moir\'e superlattice. Each eigenstate of the Hamiltonian (\ref{onesector}) can be labeled by a two-dimensional momentum $(K_1,K_2)$.

{\em Band isolation.} 
One of the most attractive advantages of TDBG is the high tunability of low-energy bands via the twist angle $\theta$ and the vertical voltage bias $U$. In particular, the first conduction band can be easily isolated and shows a rich phase diagram of band topology when $\theta$ and $U$ are varied~\cite{ashvinTDBG}. Before studying the interaction driven physics, we first explore the $U-\theta$ space to find the regions where the first conduction band is well isolated such that the band topology is well defined and the projected Hamiltonian (\ref{onesector}) may apply. 

We calculate the {\em indirect} gap $\Delta\equiv\min(\Delta_{e},\Delta_{v})$ of the first conduction band, where $\Delta_{e}$ and $\Delta_{v}$ are the indirect gaps to the next excited band and the valence band, respectively. We observe three main regions with significant $\Delta$ in the $U-\theta$ space [Fig.~\ref{TDBG_gap}(a)], in which the isolated first conduction band has Chern number $\mathcal{C}=0,1$ and $2$, respectively (similar results were reported in Ref.~\cite{ashvinTDBG} for different model parameters). Here we define $\mathcal{C}$ =  $\int_{\rm MBZ} d {\bf k} \> \Omega_C (\mathbf{k}) / (2\pi) $ with $\Omega_C(\mathbf{k}) =i\left( \braket{\frac{\partial \mu(\mathbf{k})}{\partial k_x}|\frac{\partial \mu(\mathbf{k})}{\partial k_y}} - \braket{\frac{\partial \mu(\mathbf{k})}{\partial k_y}|\frac{\partial \mu(\mathbf{k})}{\partial k_x}}\right) $ the Berry curvature of the first conduction band and $\ket{\mu(\mathbf{k})}$ the corresponding band Bloch eigenfunction. In addition, we have checked that the interaction with the valence band does not change this Chern number in the regime of parameters that we are considering~\cite{supple}. In the following, we will focus on the $\mathcal{C}=1$ and $\mathcal{C}=2$ regions in Fig.~\ref{TDBG_gap}(a), and probe the existence of robust FCIs in the first conduction band in both regions. 

{\em FCIs in the $\mathcal{C}=1$ region.}
Now we address the possibility of FCIs stabilized by the screened Coulomb interaction in the $\mathcal{C}=1$ region. We first examine $\nu=1/3$, where one may in general expect robust FCIs as the lattice analogs of the celebrated Laughlin state \cite{Laughlin}. Indeed, we observe clear three-fold ground-state degeneracies in the energy spectrum of the Hamiltonian (\ref{onesector}) for $U=20{\rm meV}-30{\rm meV}$ and $\theta=0.6^\circ-0.8^\circ$ [Fig.~\ref{C_1_nu_1_3}(a)], where the three approximately degenerate ground states have momenta $(K_1,K_2)$ which are consistent with the prediction of Haldane statistics for the $\nu=1/3$ FCIs in a $\mathcal{C}=1$ band~\cite{fciprx,manybodysymmetry,haldanestatistics}. The three-fold ground-state degeneracy persists during the magnetic flux insertion through the handles of the toroidal system, thus further confirming the robustness of finite-size fingerprint of a topological degeneracy [Fig.~\ref{C_1_nu_1_3}(b)]. The energy gap separating the three ground states from excited states becomes much larger than the ground-state splitting as the system size grows, and is very likely to survive in the thermodynamic limit as suggested by a finite-size scaling [Fig.~\ref{C_1_nu_1_3}(c)]. Remarkably, the gap corresponds to a temperature of about $10$ Kelvin, which is at least an order of magnitude higher than required by the conventional FQH states in two-dimensional electron gases. Importantly this energy scale is still smaller than the band gap, $\Delta$, confirming the validity of the band projection. 

The entanglement spectroscopy of the ground-state manifold with the particle-cut entanglement spectrum (PES)~\cite{fciprx,LiH,PES} further corroborate that the ground states are topologically nontrivial. By dividing the whole system into $N_A$ and $N-N_A$ electrons and labeling each PES level by the total momentum $(K_1^A,K_2^A)$ of those $N_A$ electrons, we find a clear entanglement gap $\Delta\xi\approx0.3$ separating the low-lying PES levels from higher ones [Fig.~\ref{C_1_nu_1_3}(d)]. The number of levels below this gap exactly matches the pertinent counting of quasihole excitations in the $\nu=1/3$ Abelian FCIs~\cite{fciprx,manybodysymmetry,haldanestatistics}, which rules out competing possibilities such as charge density waves. 

We have also considered other filling factors. At $\nu=2/3$, we find the particle-hole conjugate of the $\nu=1/3$ FCIs reported above~\cite{supple}. Moreover, we observe vestiges of a possible five-fold ground-state degeneracies at $\nu=2/5$ and $\nu=3/5$~\cite{supple}, which may suggest the $\nu=2/5$ Jain state and its particle-hole conjugate. However, these states are fragile against band dispersion and compete with spinful states (as discussed below).

{\em FCIs in the $\mathcal{C}=2$ region.}  Previous studies have reported novel FCIs residing in flat bands with higher Chern numbers~\cite{highCWang,highCLiu,max,highCYang,highCSterdyniak,highCMoller,highCWu}. Unlike FCIs in $|\mathcal{C}|=1$ flat bands, these high-$\mathcal{C}$ FCIs do not have usual continuum FQH states as counterparts. In particular, application of composite fermion theory to Bloch bands predicts a series of high-$\mathcal{C}$ FCIs at $\nu=r/(rk|\mathcal{C}|+1)$~\cite{highCMoller}, where $k>0$ is the number of flux attached to each particle in the composite fermion theory ($k$ should be even for fermions) and $r$ is the number of fully filled composite-fermion bands. Note that $r$ can be negative which corresponds to negative flux attachment. The energy gap usually decays with the decreasing of $\nu$---specifically with increasing denominators---which makes the states more fragile and harder to observe both in finite-size calculations and experiments. Consequently, we set $k=2$ and $r=-1$ in the following to consider the largest filling $\nu=1/3$ in the $\nu=r/(2rk+1)$ branch for the $\mathcal{C}=2$ region of TDBG. 

Remarkably, we find clear three-fold ground-state degeneracies for relatively large systems ($N\geq8$ electrons)  with $U=50{\rm meV}-70{\rm meV}$ and $\theta=1.2^\circ-1.4^\circ$ [Fig.~\ref{C_2_nu_1_3}(a)]. While the energies vary more during the flux insertion than at $\nu=1/3$ in the $\mathcal{C}=1$ region, the three-fold topological degeneracy still persists in the spectral flow [Fig.~\ref{C_2_nu_1_3}(b)]. The finite-size scaling for the three available system sizes within our computational limit suggests that an energy gap of $\sim5$ Kelvin probably survives in the thermodynamic limit [Fig.~\ref{C_2_nu_1_3}(c)]. Moreover, we obtain similar results~\cite{supple} also in the exactly flat $\mathcal{C}=2$ band in the chiral limit of TDBG~\cite{alexnanoletters}, where the FCIs at $\nu=1/3$ should be topologically equivalent to what we observe here in the realistic TDBG model. In that case, we further observe a clear entanglement gap in the PES, which unambiguously rules out the charge density wave and suggests a nontrivial topological order distinct from the $\nu=1/3$ Laughlin state~\cite{supple}. All these results confirm the topological nontrivial property of the $\nu=1/3$ states in the $\mathcal{C}=2$ region of TDBG.


\begin{figure}
	\centerline{\includegraphics[width=0.9\linewidth]{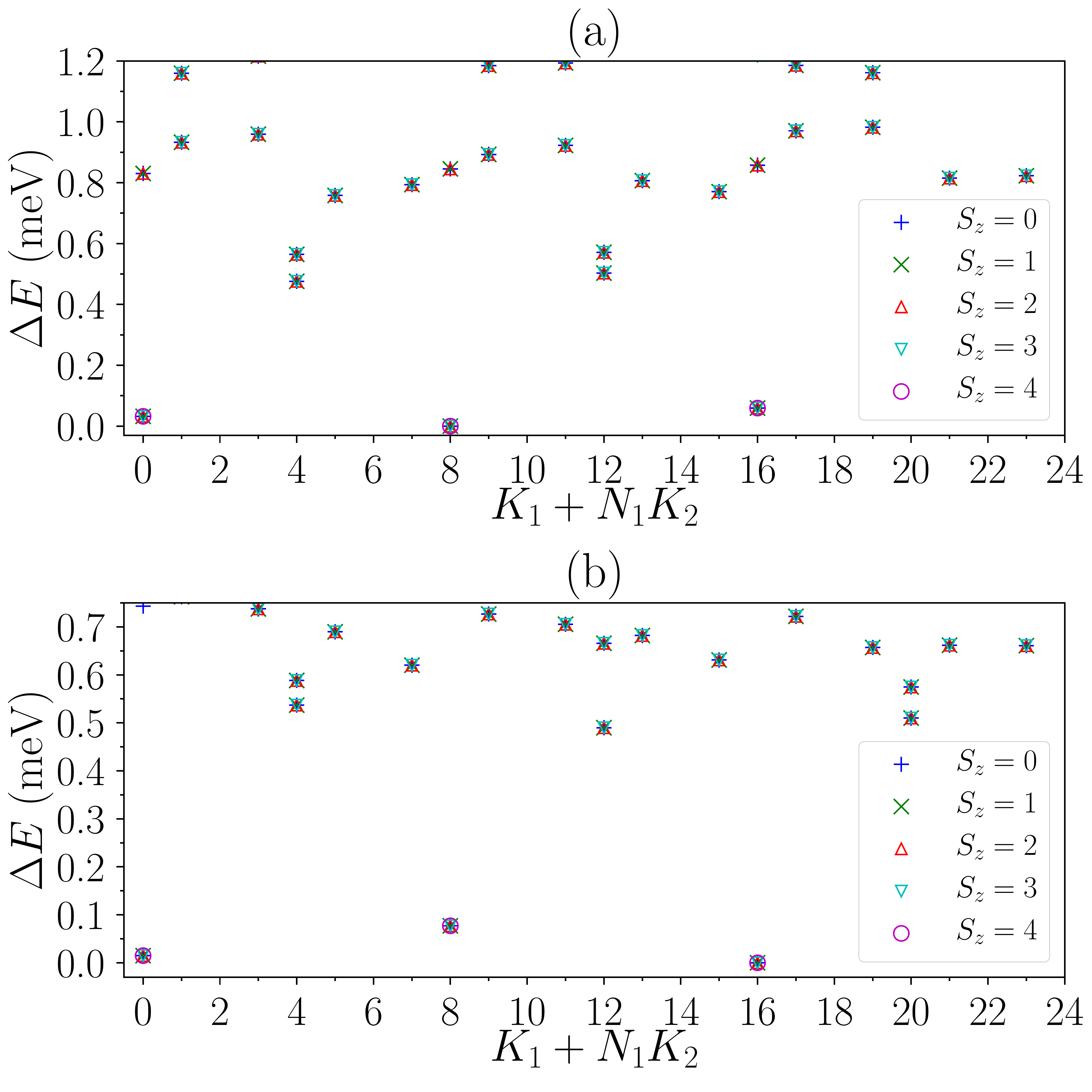}}
	\caption{The low-lying energy spectra at $\nu=1/3$ for $N=8$ spinful electrons on the $N_1\times N_2=4\times6$ lattice. The Chern number of the first conduction band is $\mathcal{C}=1$ in (a) and $\mathcal{C}=2$ in (b), with $\theta=0.75^\circ$, $U=20{\rm meV}$ and $\theta=1.35^\circ$, $U=60{\rm meV}$, respectively.}
	\label{1_3_spin}
\end{figure}

{\em Spinful FCIs.} So far we have assumed both valley and spin polarization. Now motivated by experiments we keep valley polarization \cite{correlatedTDBG2}, but bring the spin degree of freedom back. In this case, as both the band dispersion and the interaction are independent on the spin flavor if one neglects Hund's coupling terms that are generically much weaker than the Coulomb interaction, the spinful many-body Hamiltonian~\cite{supple} is SU(2) symmetric within each valley, which allows us to label each energy level by the total spin $S$ and its $z$-component $S_z=(N_\uparrow-N_\downarrow)/2$, with $N_\uparrow$ and  $N_\uparrow$ the number of spin-up and spin-down electrons, respectively. At $\nu=1/3$ in both the $\mathcal{C}=1$ and $\mathcal{C}=2$ regions, we find three-fold ground-state degeneracies in all $S_z$ sectors and the ground energies with different $S_z$ are identical (Fig.~\ref{1_3_spin}). This confirms the assertion that the $\nu=1/3 $ FCIs observed in both $\mathcal{C}=1$ and $\mathcal{C}=2$ regions are indeed ferromagnetic with total spin $S=N/2$. 

\begin{figure}
	\centerline{\includegraphics[width=0.9\linewidth]{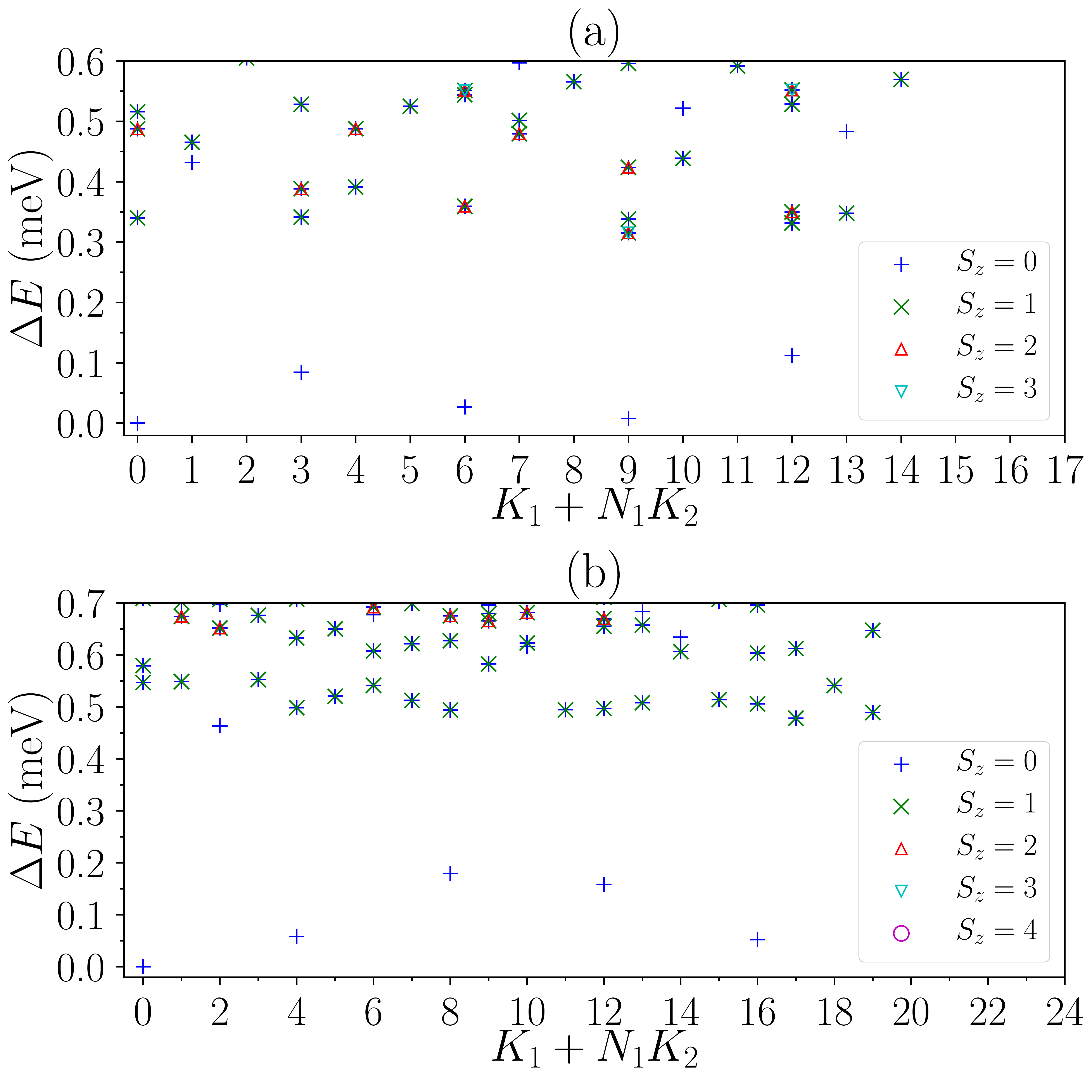}}
	\caption{The low-lying energy spectra at $\nu=2/5$ for (a) $N=6$ spinful electrons on the $N_1\times N_2=3\times5$ lattice and (b) $N=8$ spinful electrons on the $N_1\times N_2=4\times5$ lattice. Here we choose $\theta=0.65^\circ$, $U=20{\rm meV}$.}
	\label{332}
\end{figure}

On the other hand, we find that such ferromagnetism can disappear at other filling factors, leading to spin-singlet ground states. For example, at $\nu=2/5$ in the $\mathcal{C}=1$ region with $U$ around $20{\rm meV}$ and $\theta=0.65^\circ-0.75^\circ$, the ground states are in the $S_z=0$ sector, thus having total $S=0$. Remarkably, five-fold ground-state degeneracies appear in this case (Fig.~\ref{332}), strongly suggesting the Halperin $(332)$ state as the ground state. As it was found in other graphene based Moir\'e systems that ferromagnetic states are favored with the increasing of $w_0/w_1$~\cite{CecileFCI}, there could be a phase transition from the Halperin $(332)$ state to the $\nu=2/5$ Jain state at $w_0/w_1$ larger than the value $0.7$ chosen in this work.

{\em Discussion.} In this work we have established twisted double bilayer graphene as a flexible platform for a plethora of fractional Chern insulators which are stabilized by tuning a gate voltage and the twist angle. Remarkably, these exotic states occur at high temperatures and some of them occur in parameter regimes that are directly experimentally accessible. In particular, the novel $\mathcal C=2$ FCIs at filling $\nu=1/3$ which we predict around a range of twist angles $\theta \approx 1.2^\circ - 1.4^\circ$ are perfectly within the reach of current experiments, in which TDBG samples with $\theta\approx 0.84^\circ - 2^\circ$ have been manufactured~\cite{correlatedTDBG1,correlatedTDBG2,pablotdbgnature,correlatedTDBG3}. While it remains challenging in experiments to realize TDBG samples with $\theta$ as low as $ 0.6^\circ - 0.8^\circ$ where we numerically observe the $\mathcal{C} = 1$ FCIs, future advances in manafacturing van der Waals heterostructures might make this possible. Prospective experimentally realizable FCIs indeed provide an intriguing path toward the possibility of many technological applications. In the latter context, the investigation of possible non-Abelian states provides a particularly intriguing outlook. 

\acknowledgments
{\it Acknowledgments.}
We would like to thank Alex Kruchkov, Jose Lado and Yoran Tournois for useful discussions. A.~A. and E.~J.~B. are supported by the Swedish Research Council (VR) and the Wallenberg Academy Fellows program of the Knut and Alice Wallenberg Foundation. Z.~L. is supported by the National Natural Science Foundation of China through Grant No.~11974014.

\newpage
\onecolumngrid 
\newpage

\section{Supplementary Material}
\setcounter{subsection}{0}
\setcounter{equation}{0}
\setcounter{figure}{0}
\renewcommand{\theequation}{S\arabic{equation}}
\renewcommand{\thefigure}{S\arabic{figure}}
\renewcommand{\thesubsection}{S\arabic{subsection}}

In this supplementary material, we provide details about the single-particle model of twisted double bilayer graphene (TDBG) and the projected interaction, present more numerical results for FCIs in the $\mathcal{C}=1$ region, and discuss the effects of interaction, screening mechanism, and parameter choice on our results.

\maketitle
\subsection{TDBG model}
We consider two sheets of AB stacked bilayer graphene (BLG) which are twisted with respect to one another, with the ABAB stacking pattern. Each AB stacked BLG sheet is modeled by the following single-particle Hamiltonian 
\begin{eqnarray}
h({\bf k})=
\left(
\begin{array}{cccc}
U_1+\delta&\frac{\sqrt{3}}{2}at_0(k_x-ik_y)&-\frac{\sqrt{3}}{2}at_4(k_x+ik_y)&t_1\\
\frac{\sqrt{3}}{2}at_0(k_x+ik_y)&U_1&-\frac{\sqrt{3}}{2}at_3(k_x-ik_y)&-\frac{\sqrt{3}}{2}at_4(k_x+ik_y)\\
-\frac{\sqrt{3}}{2}at_4(k_x-ik_y)&-\frac{\sqrt{3}}{2}at_3(k_x+ik_y)&U_2&\frac{\sqrt{3}}{2}at_0(k_x-ik_y)\\
t_1&-\frac{\sqrt{3}}{2}at_4(k_x-ik_y)&\frac{\sqrt{3}}{2}at_0(k_x+ik_y)&U_2+\delta
\end{array}\right)
\label{smeq1}
\end{eqnarray}
near the valley $\mathbf{K}_{+} = \frac{4\pi}{3 a}(1,0) $ of its Brillouin zone, where $a\approx 2.46{\rm \AA}$ is the lattice constant of graphene, and the basis is $(\psi_{A_1}({\bf k}),\psi_{B_1}({\bf k}),\psi_{A_2}({\bf k}),\psi_{B_2}({\bf k}))^T$ with $A,B$ the sublattice indices of monolayer graphene and $1,2$ the layer indices in the AB stacked BLG sheet. $t_0$, $t_1$, $t_3$ and $t_4$ are the $A_1-B_1$, $A_1-B_2$, $B_1-A_2$ and $B_1-B_2$ hopping strengths in the AB stacked BLG sheet, respectively, $\delta$ is the onsite energy difference between $A$ and $B$ sites, and $U_i$'s are the gating voltage across the system. Throughout the work we adopt $(t_0,t_1,t_3,t_4,\delta)=(2610,361,283,138,15){\rm meV}$~\cite{ABstacked,ashvinTDBG}.  

When the two AB stacked BLG sheets are twisted, we focus on the Moir\'e Brillouin zone of TDBG formed near $\mathbf{K}_{+}$. The two primitive reciprocal lattice vectors of TDBG are chosen as ${\bf G}_1=\frac{2\pi}{a_M}(\frac{1}{\sqrt{3}},1)$ and ${\bf G}_1=\frac{2\pi}{a_M}(-\frac{1}{\sqrt{3}},1)$, with $\theta$ the twist angle and $a_M=a/(2\sin\frac{\theta}{2})$ the lattice constant of TDBG. Let us denote ${\bf K}_+^t=R_{\theta/2}{\bf K}_+$ and ${\bf K}_+^b=R_{-\theta/2}{\bf K}_+$, where $R_{\theta} $ a counter-clockwise rotation around the $z$-axis in the momentum space. The single-particle Hamiltonian of TDBG for each spin favor can then be written as 
\begin{eqnarray}
H =  \sum_{\mathbf{k}} \psi^{\dagger}_t(\mathbf{k}) h_{-\theta/2}(\mathbf{k}-{\bf K}_+^t) \psi_{t}(\mathbf{k}) + \sum_{\mathbf{k}} \psi^{\dagger}_b(\mathbf{k}) h_{\theta/2}(\mathbf{k}-{\bf K}_+^b) \psi_{b}(\mathbf{k}) + \sum_{\mathbf{k}}\sum_{ j = 0}^2 \big(\psi_{t}^{\dagger}(\mathbf{k}-{\bf q}_0+\mathbf{q}_j) T_j \psi_{b}(\mathbf{k}) + h.c. \big),  \nonumber\\
\label{smeq2}
\end{eqnarray}
where $ \psi_{l=t,b}({\bf k}) = (\psi_{A_{1,l}}({\bf k}),\psi_{B_{1,l}}({\bf k}),\psi_{A_{2,l}}({\bf k}),\psi_{B_{2,l}}({\bf k}))^T$ is the basis for electrons in top and bottom BLG sheets, respectively, $h_{\theta}(\mathbf{k}) = h(R_{\theta} \mathbf{k}) $ with $h(\mathbf{k})$ given in Eq.~(\ref{smeq1}), and $\mathbf{q}_0 = R_{-\theta/2} \mathbf{K}_+ - R_{\theta/2} \mathbf{K}_+$, $\mathbf{q}_1  = R_{2\pi/3} \mathbf{q}_0$ and $\mathbf{q}_2 = R_{-2\pi/3} \mathbf{q}_0$.
We choose as $U_1=U/2, U_2=U/6$ for the top BLG sheet and $U_1=-U/6, U_2=-U/2$ for the bottom BLG sheet. The Moir\'e hoppings $T_j$'s in TDBG are given by
\begin{equation}
T_j = w_0 + w_1 e^{i(2\pi/3)j \sigma_z} \sigma_x e^{-i(2\pi/3)j\sigma_z},
\end{equation} 
where $w_0$ and $w_1$ are the inter-sheet hopping strengths between $AA$ and $AB$ sites, respectively. In our numerics, we take $ w_1 = 100 $ meV and $w_0 = 0.7 w_1$. 

For each ${\bf k}_0$ in the MBZ, by writing ${\bf k}$ in the single-particle Hamiltonian $H$ [Eq.~(\ref{smeq2})] as ${\bf k}_0+m{\bf G}_1+n{\bf G}_2$ and setting integers $m,n=-d,...,d$, we can construct $H({\bf k}_0)$ as a matrix of dimension $8(2d+1)^2$. The eigenvalues and eigenvectors of this Hamiltonian matrix then give us the band energies and eigenvectors of TDBG at ${\bf k}_0$. We choose $d=7$ in our calculations to reach convergence of numerical results. 

\subsection{Projected many-body Hamiltonian}
Now we give the matrix elements $V_{\mathbf{k}_1\mathbf{k}_2\mathbf{k}_3\mathbf{k}_4}$ of the two-body interaction projected to the first conduction band of TDBG. For specific spin and valley,  we have
\begin{eqnarray}
\label{projham}
V_{\mathbf{k}_1\mathbf{k}_2\mathbf{k}_3\mathbf{k}_4} &=& \frac{1}{2} \sum_{\mathbf{q} } V(\mathbf{q}) \sum_{s,s'} \sum_{\{m_i,n_i\} = -d}^{d}  \delta_{\mathbf{k}_1 + \mathbf{k}_2 + (m_1+m_2)\mathbf{G}_1 + (n_1+n_2)\mathbf{G}_2, \mathbf{k}_3 + \mathbf{k}_4 + (m_3+m_4)\mathbf{G}_1 + (n_3+n_4)\mathbf{G}_2}   \nonumber\\ 
& \times&   \delta_{\mathbf{k}_1 - \mathbf{k}_4 + (m_1-m_4)\mathbf{G}_1 + (n_1 - n_4)\mathbf{G}_2,\mathbf{q}}  \mu^{*}_{m_1,n_1,s}(\mathbf{k}_1) \mu^{*}_{m_2,n_2,s'}(\mathbf{k}_2) \mu_{m_3,n_3,s'}(\mathbf{k}_3)\mu_{m_4,n_4,s}(\mathbf{k}_4).
\end{eqnarray}
Here $V({\bf q})$ is the Fourier transform of the interaction potential. For the Yukawa potential, $V({\bf q})=\frac{e^2}{4\pi\epsilon S}\frac{2\pi}{\sqrt{|{\bf q}|^2+\kappa^2}}$, where $e$ is the electron charge, $\epsilon$ is the dielectric constant of the material, $S$ is the area of the Moir\'e superlattice, and $\kappa$ measures the screening strength. $ \{\mu_{m,n,s}(\mathbf{k}_i) \}$ is the eigenvector of the first conduction band obtained from diagonalizing the single-particle Hamiltonian Eq.~(\ref{smeq2}) for ${\bf k}_i\in{\rm MBZ}$, with ${\bf k}={\bf k}_i+m{\bf G}_1+n{\bf G}_2$ in Eq.~(\ref{smeq2}) and $s=(A/B)_{(1/2),(t/b)}$ the orbital index. For a finite periodic system with $N_1\times N_2$ unit cells, where $N_1$ and $N_2$ are the number of unit cells in the two basic directions ${\bf a}_1=a_M(\frac{\sqrt{3}}{2},\frac{1}{2})$ and ${\bf a}_2=a_M(-\frac{\sqrt{3}}{2},\frac{1}{2})$ of TDBG, ${\bf k}_i$ takes the value $\frac{k_i^1}{N_1}{\bf G}_1+\frac{k_i^2}{N_2}{\bf G}_2$ with $k_i^1=0,1,\cdots,N_1-1$ and $k_i^2=0,1,\cdots,N_2-1$. 

As the matrix elements of the single-particle Hamiltonian Eq.~(\ref{smeq2}) are identical for spin-up and spin-down electrons, we can generalize the projected many-body total Hamiltonian [Eq.~(1) in the main text] to the spinful case: 
\begin{eqnarray}
H^{\rm proj} = \sum_{{\bf k}.\sigma}E({\bf k}) c_{{\bf k}.\sigma}^\dagger c_{{\bf k},\sigma}+\sum_{\{\mathbf{k}_i\}}\sum_{\sigma,\sigma'} V_{\mathbf{k}_1\mathbf{k}_2\mathbf{k}_3\mathbf{k}_4} c^\dagger_{{\mathbf{k}_1},\sigma} c^\dagger_{{\mathbf{k}_2},\sigma'} c_{{\mathbf{k}_3},\sigma'} c_{{\mathbf{k}_4},\sigma},
\end{eqnarray} 
where $c^\dagger_{{\mathbf{k}},\sigma}$ ($c_{{\mathbf{k}},\sigma}$) creates (annihilates) an electron with momentum $\mathbf{k}$ and spin $\sigma$ in the first conduction band (per valley). Note that both the band dispersion $E({\bf k})$ and the interaction matrix element $V_{\mathbf{k}_1\mathbf{k}_2\mathbf{k}_3\mathbf{k}_4}$ are independent on the spin, leading to an SU(2) symmetric Hamiltonian. 
\begin{figure*}
\centerline{\includegraphics[width=\linewidth]{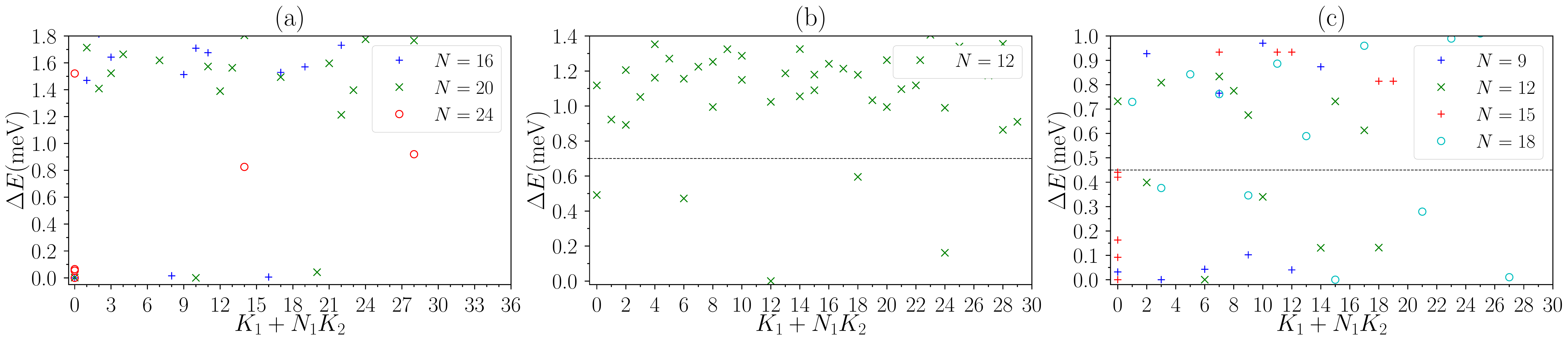}}
\caption{(a) The low-lying energy spectra at $\nu=2/3$ for $N=16, N_1\times N_2=4\times6$, $N=20,N_1\times N_2=5\times6$ and $N=24,N_1\times N_2=6\times6$, with $\theta=0.75^\circ$, $U=20{\rm meV}$. (b) The low-lying energy spectra at $\nu=2/5$ for $N=12, N_1\times N_2=6\times5$, with $\theta=0.75^\circ$, $U=20{\rm meV}$. (c) The low-energy spectra at $\nu=3/5$ for $N=9, N_1\times N_2=3\times5$, $N=12, N_1\times N_2=4\times5$, $N=15, N_1\times N_2=5\times5$ and $N=18, N_1\times N_2=6\times5$, with $\theta=0.8^\circ$, $U=20{\rm meV}$. The dashed lines in (b) and (c) separate the five ground states from excited levels.}
\label{C_1_more}
\end{figure*}
\subsection{$\nu=2/3$, $\nu=2/5$ and $\nu=3/5$ in the $\mathcal{C}=1$ region}
Here we assume both spin and valley polarization. We first examine $\nu=2/3$ in the $\mathcal{C}=1$ region. With the same parameters as where we find the $\nu=1/3$ FCIs, we again observe nice three-fold ground-state degeneracies at $\nu=2/3$, as shown in Fig.~\ref{C_1_more}(a). We identify these  states as the particle-hole conjugates of the $\nu=1/3$ FCIs in the $\mathcal{C}=1$ region shown in the main text.   

We then report the numerical results at $\nu=2/5$ and $\nu=3/5$ in the $\mathcal{C}=1$ region of TDBG. In both cases, our numerical results do not suggest well developed FCIs for finite systems within our computational limit and the model parameters which we choose. However, we still observe five lowest energy levels in the correct momentum sectors predicted for the $\nu=2/5$ Jain FCIs and their $\nu=3/5$ particle-hole conjugates, although the splitting between these states are significantly larger than the $\nu=1/3$ cases shown in the main text [Figs.~\ref{C_1_more}(b) and \ref{C_1_more}(c)]. Therefore, the $\nu=2/5$ Jain FCIs and their $\nu=3/5$ particle-hole conjugates could be stabilized for larger system sizes or modified model parameters (especially $w_0$ and $w_1$).

\subsection{Dependence of the Chern number on the interaction}
Let us first clarify the definition of band Chern number that we adopt in this work. Chern number characterizes the topology of an isolated band in two-dimensional lattice systems. In the most general case, it is related to the Hall conductance $\sigma_{xy}$ of the band by $\sigma_{xy}=\frac{e^2}{h}\mathcal{C}$, where $h$ is Planck constant. According to the Kubo formula, we have
\begin{equation} 
\mathcal{C}=i\frac{2\pi\hbar^2}{e^2S}\sum_{m\in s}\sum_{n\notin s}\frac{\langle m|J_x|n\rangle\langle n|J_y|m\rangle-\langle m|J_y|n\rangle\langle n|J_x|m\rangle}{(E_m-E_n)^2},
\label{chernnumber1}
\end{equation} 
where $s$ labels the band, $S$ is the lattice area, ${\bf J}$ is the current operator, and $E_m$ and $|m\rangle$ are eigenvalue and eigenstate of the single-particle Hamiltonian $H$, respectively. Evaluating this quantity may require the information of the density of states in the energy representation.

In the absence of disorder (like in our case), as each eigenvalue and eigenstate of the single-particle Hamiltonian can be labeled by a momentum ${\bf k}$, one can transform the sum over single-particle states in Eq.~(\ref{chernnumber1}) to a sum over band index and another sum (integral in the thermodynamic limit) over momentum ${\bf k}$, which gives~\cite{Thouless}
\begin{equation} 
\mathcal{C}=\frac 1 {2\pi}\int_{\rm{BZ}} \Omega_s({\bf k})\mathrm{d}^2{\bf k},
\label{chernnumber}
\end{equation} 
with
\begin{eqnarray}
\Omega_{s}({\bf k})=i\sum_{p\neq s}\left[\frac{\langle \psi_{s}({\bf k})|\partial_{k_x} H({\bf k})|\psi_{p}({\bf k})\rangle\langle \psi_{p}({\bf k})|\partial_{k_y} H({\bf k})|\psi_{s}({\bf k})\rangle}{[E_{s}({\bf k})-E_{p}({\bf k})]^2}-{\rm h.c.}\right]
\label{curvature1}
\end{eqnarray} 
the Berry curvature of band $s$. Here $H({\bf k})$ is the single-electron Hamiltonian in the $k$-space, $|\psi_p({\bf k})\rangle$ is the eigenvector of band $p$, $E_p({\bf k})$ is the energy of band $p$, the integral in Eq.~(\ref{chernnumber}) is over the primitive Brillouin zone (BZ), and the sum in Eq.~(\ref{curvature1}) is over all bands except $s$. Although the Chern number Eq.~(\ref{chernnumber}) seemingly depends on the band energies which appear in Eq.~(\ref{curvature1}), we can use the relation $\langle\psi_p({\bf k})|\partial_{k_{x,y}}\psi_s({\bf k})\rangle=\frac{\langle \psi_{p}({\bf k})|\partial_{k_{x,y}} H({\bf k})|\psi_{s}({\bf k})\rangle}{E_{s}({\bf k})-E_{p}({\bf k})}$ for $p\neq s$ to remove band energies and the Hamiltonian from the expression of Berry curvature, leading to
\begin{eqnarray}
\Omega_s({\mathbf k})
=i\left(\langle{\partial_{k_x} \psi_s(\mathbf{k})|\partial_{k_y} \psi_s(\mathbf{k})}\rangle - \langle{\partial_{k_y} \psi_s(\mathbf{k})|\partial_{k_x} \psi_s(\mathbf{k})}\rangle\right)
\label{berrycurvature}
\end{eqnarray}
which only contains the eigenvector of band $s$. Now it is clear that the band Berry curvature and Chern number $\mathcal{C}$ depend only on the eigenstate of the pertinent band and are independent of the energy dispersion. Moreover, even if the band eigenvector is modified such that the Berry curvature Eq.~(\ref{berrycurvature}) changes, the integral of Berry curvature over the BZ -- $\mathcal{C}$, is an integer valued topological invariant in the sense that it cannot change so long as band $s$ keeps isolated from other bands. 

For a finite lattice with $N_1\times N_2$ unit cells, we use the method proposed in Ref.~\cite{chern} to numerically evaluate Eq.~(\ref{chernnumber}). In this case, the Brillouin zone is discrete with $N_1N_2$ allowed ${\bf k}$ points, where ${\bf k}=\frac{k_1}{N_1}{\bf g}_1+\frac{k_2}{N_2}{\bf g}_2$, $k_1=0,\cdots,N_1-1$ and $k_2=0,\cdots,N_2-1$ are integers and ${\bf g}_1$ and ${\bf g}_2$ are primitive reciprocal lattice vectors. For each ${\bf k}$ point, we define a U(1) link 
\begin{equation} 
U_\alpha({\bf k})=\frac{\langle\psi_s({\bf k})|\psi_s({\bf k}+{\bf g}_\alpha/N_\alpha)\rangle}{|\langle\psi_s({\bf k})|\psi_s({\bf k}+{\bf g}_\alpha/N_\alpha)\rangle|}
\end{equation} 
with $\alpha=1,2$, such that the integral of Berry curvature over the parallelogram with vertices ${\bf k}$, ${\bf k}+{\bf g}_1/N_1$, ${\bf k}+{\bf g}_2/N_2$ and ${\bf k}+{\bf g}_1/N_1+{\bf g}_2/N_2$ can be approximated by 
\begin{equation} 
F({\bf k})=\Im \ln[U_1({\bf k})U_2({\bf k}+{\bf g}_1/N_1)U_1^*({\bf k}+{\bf g}_2/N_2)U_2^*({\bf k})],
\end{equation} 
where $\Im$ means imaginary part. This approximately gives the Chern number as $\mathcal{C}=\frac{1}{2\pi}\sum_{\bf k} F({\bf k})$. Then we increase $N_1$ and $N_2$ to reach the value in the thermodynamic limit. 

Now we would like to investigate whether the interaction can change the Chern number of the first conduction band in the TDBG model that we adopt. To that end, we consider the interaction between the first conduction band and the valence band below. We neglect effects from bands lower than the valence band and assume that bands above the first conduction band are empty. So we have a system consisting of a completely filled valence band and a partially filled conduction band at a fixed filling $\nu$. The Hamiltonian projected into the subspace of these two bands is given by 
\begin{equation}
H = \sum_{\mathbf{k},\alpha} E_\alpha(\mathbf{k}) c^\dagger_\alpha(\mathbf{k})c_\alpha(\mathbf{k}) + \frac{1}{2} \sum_{\mathbf{q}} V(\mathbf{q}) :\tilde{\rho}(\mathbf{q}) \tilde{\rho}(-\mathbf{q}): 
\label{eqs6}
\end{equation} 
with the projected density operator 
\begin{equation}
\tilde{\rho}(\mathbf{q}) = \sum_{\mathbf{k},\alpha,\beta} \lambda^{\alpha\beta}(\mathbf{k}+\mathbf{q},\mathbf{k})c^\dagger_\alpha(\mathbf{k}+\mathbf{q})c_\beta(\mathbf{k}),
\end{equation} 
where $c^\dagger_\alpha(\mathbf{k})$ creates an electron with momentum $\mathbf{k}$ in band $\alpha = 1,2$, corresponding to the valence and conduction band, respectively, and $E_{\alpha}(\mathbf{k})$ is the bare band dispersion. The form factor $\lambda^{\alpha\beta}(\mathbf{k}_1,\mathbf{k}_2)$ is defined as \begin{equation}
\lambda^{\alpha\beta}(\mathbf{k}_1,\mathbf{k}_2) = \braket{\mu^\alpha(\mathbf{k}_1)|\mu^\beta(\mathbf{k}_2)},
\end{equation} 
with $|\mu^\alpha(\mathbf{k}_i)\rangle$ is the eigenvector of band $\alpha$ obtained from diagonalizing the single-particle Hamiltonian with ${\bf k}={\bf k}_i+m{\bf G}_1+n{\bf G}_2$ in Eq.~(\ref{smeq2}). We choose a periodic gauge $\mu^\alpha_{m,n,a}(\mathbf{k}_0 + m_0 \mathbf{G}_1 + n_0 \mathbf{G}_2) = \mu^\alpha_{m+m_0,n+n_0,a}(\mathbf{k}_0) $ for $\mathbf{k}_0 \in \text{MBZ}$ to project the momenta in the arguments of $c_\alpha(\mathbf{k})$ and $\lambda^{\alpha\beta}(\mathbf{k}_1,\mathbf{k}_2)$ into the MBZ. Then Eq.~(\ref{eqs6}) is basically the same as Eq.~\eqref{projham}, but now we project the Hamiltonian to both the conduction band and the valence band and we also fix the gauge choice. 

\begin{figure*}[t]
\centerline{\includegraphics[width=\linewidth]{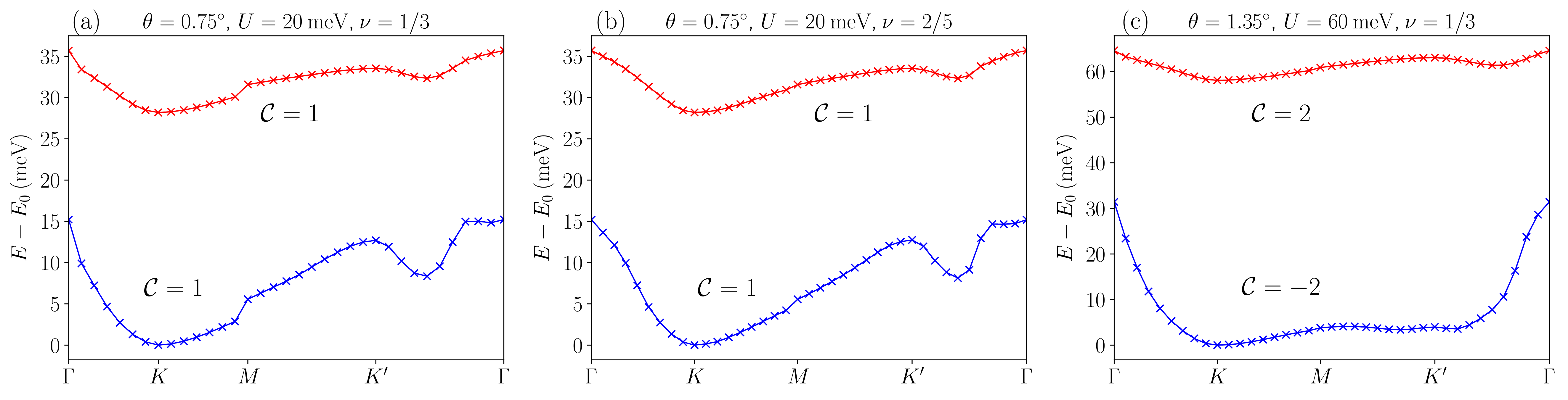}}
\caption{Renormalized valence (conduction) band structure shown in blue (red)  with their associated Chern number $\mathcal{C}$ for selective values of $\theta$ and $U$ for different fillings $\nu$ of the conduction band with the valence band fully filled. Energies are plotted relative to the minimum valence band energy $E_0$ along $\Gamma=(0,0)$, $K = K_M(\sqrt{3}/2,1/2) $, $M = K_M(\sqrt{3}/4,3/4)$, and $K' = K_M (0,1)$ with $K_M = \frac{4\pi}{3 a_M}$. The relative dielectric constant $\epsilon_r$ is set to $\epsilon_r = 4$.}
\label{renormalizedbandstructure}
\end{figure*}

Next we do the standard Hartree Fock (HF) mean-field procedure by replacing bilinear operators in Eq.~(\ref{eqs6}) with their expectation values, i.e., \begin{gather*}
c^{\dagger}_\alpha(\mathbf{k}_1+\mathbf{q}) c^{\dagger}_\gamma(\mathbf{k}_2-\mathbf{q}) c_\delta(\mathbf{k}_2)c_\beta(\mathbf{k}_1) = \\ c^{\dagger}_\alpha(\mathbf{k}_1+\mathbf{q}) c_\beta(\mathbf{k}_1) \braket{c^{\dagger}_\gamma(\mathbf{k}_2-\mathbf{q}) c_\delta(\mathbf{k}_2)} + c^{\dagger}_\gamma(\mathbf{k}_2-\mathbf{q}) c_\delta(\mathbf{k}_2) \braket{c^{\dagger}_\alpha(\mathbf{k}_1+\mathbf{q}) c_\beta(\mathbf{k}_1)} \\ - c^{\dagger}_\alpha(\mathbf{k}_1+\mathbf{q})  c_\delta(\mathbf{k}_2) \braket{c^{\dagger}_\gamma(\mathbf{k}_2-\mathbf{q})c_\beta(\mathbf{k}_1)} - c^{\dagger}_\gamma(\mathbf{k}_2-\mathbf{q})  c_\beta(\mathbf{k}_1) \braket{c^{\dagger}_\alpha(\mathbf{k}_1+\mathbf{q})c_\delta(\mathbf{k}_2)}.
\end{gather*}
Given the conserved momentum and fixed band fillings, we assume $\braket{c^\dagger_\alpha(\mathbf{k}_1)c_\beta(\mathbf{k}_2)} = \delta_{\alpha,\beta} \>  \delta_{\mathbf{k}_1,\mathbf{k}_2} \Theta(E_F-E_\alpha(\mathbf{k}_1)) $ at zero temperature, where $\Theta(x)$ is the Heaviside function and $E_F$ is the Fermi energy. For the valence band that is completely filled, we simply have $\Theta(E_F-E_\alpha(\mathbf{k}))=1$. For the conduction band at filling $\nu$, electrons will fill a subset of available states that corresponds to the lowest energies. Putting everything together, we have \begin{equation}
\label{Ham}
H = \sum_{\mathbf{k},\alpha}E_\alpha(\mathbf{k}) c^\dagger_\alpha(\mathbf{k})c_\alpha(\mathbf{k}) + \sum_{{\bf k},\alpha,\beta}\Sigma_{\alpha\beta}(\mathbf{k})c^\dagger_\alpha(\mathbf{k})c_\beta(\mathbf{k}),
\end{equation}
where the HF energy is given by \begin{equation}
\label{HF}
\begin{aligned}
\Sigma_{\alpha\beta}(\mathbf{k}) = \sum_{\mathbf{k}'}  \sum_{m,n,\gamma} \Theta(E_F-E_\gamma(\mathbf{k}')) \bigg(V(m\mathbf{G}_1+n\mathbf{G}_2)\lambda^{\alpha\beta}(\mathbf{k}+m\mathbf{G}_1+n\mathbf{G}_2,\mathbf{k})\lambda^{\gamma\gamma}(\mathbf{k}'-m\mathbf{G}_1-n\mathbf{G}_2,\mathbf{k}') \\ - V(\mathbf{k}-\mathbf{k}'+m\mathbf{G}_1+n\mathbf{G}_2)  \lambda^{\alpha\gamma}(\mathbf{k}+m\mathbf{G}_1+n\mathbf{G}_2,\mathbf{k}')\lambda^{\gamma\beta}(\mathbf{k}'-m\mathbf{G}_1-n\mathbf{G}_2,\mathbf{k})  \bigg).
\end{aligned}
\end{equation}
The above equation needs to be solved self-consistently. The right hand side depends on the form factors given by the eigenfunctions of the Hamiltonian. Starting from zero initial conditions with $\Sigma_{\alpha\beta} = 0$, we have the eigenfunctions $|\mu^{\alpha}(\mathbf{k})\rangle$ that give rise to the initial Chern numbers. For each specific ${\bf k}\in{\rm MBZ}$, $|\mu^{\alpha}(\mathbf{k})\rangle$ span a $2\times2$ subspace of the initial valence and conduction band, in which the Hamiltonian Eq.~\eqref{Ham} is represented as a $2\times2$ matrix $h({\bf k})$, whose diagonal elements are the initial bare dispersion $E_\alpha(\mathbf{k})$ and off-diagonal terms are the HF energy $\Sigma_{\alpha\beta}(\mathbf{k})$ obtained by plugging $|\mu^{\alpha}(\mathbf{k})\rangle$ in Eq.~\eqref{HF}. Then we diagonalize $h({\bf k})$ and get new band eigenfunctions that are linear combination of the initial ones:
\begin{equation}
|\tilde{\mu}^{\alpha}(\mathbf{k})\rangle = \sum_{\beta} \Gamma^{\alpha}_{\beta}(\mathbf{k}) |\mu^{\beta}(\mathbf{k})\rangle,
\end{equation} 
where $\Gamma^{\alpha}_{\beta}(\mathbf{k})$ is component $\beta$ of the $\alpha$-th eigenvector of $h({\bf k})$. We then update $h({\bf k})$ by evaluating Eq.~\eqref{Ham} in the new subspace spanned by $|\tilde{\mu}^{\alpha}(\mathbf{k})\rangle$ and keep iterating until convergence. The converged eigenfunctions are then used to compute the new Chern numbers. 

By including the interaction effect in this way, we have confirmed that the Chern number of the first conduction band is not changed by the interaction for the parameter sets we use to explore FCIs: $(\epsilon_r, \theta, U, \nu) = (4,0.75^\circ,20 \rm meV,1/3), (4,0.75^\circ,20\rm meV,2/5),(4,1.35^\circ,60 meV,1/3)$ where $\epsilon_r$ is the relative dielectric constant of the material and $\nu$ is the filling of the conduction band. The renormalized band dispersion is shown in Fig \ref{renormalizedbandstructure}.

\subsection{Dependence on the screening mechanism}
\begin{figure*}[t]
	\centerline{\includegraphics[width=\linewidth]{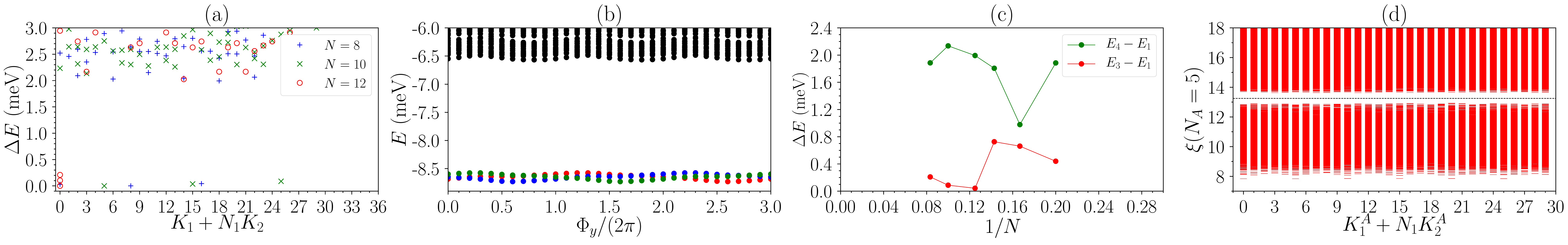}}
	\caption{Evidence of $\nu=1/3$ FCIs stabilized by the bare Coulomb interaction in the $\mathcal{C}=1$ region with $\theta=0.75^\circ$, $U=20{\rm meV}$. (a) The low-lying energy spectrum for $N=8, N_1\times N_2=4\times6$, $N=10,N_1\times N_2=5\times6$ and $N=12,N_1\times N_2=6\times6$. (b) The spectral flow for $N=10,N_1\times N_2=5\times 6$, where $\Phi_y$ is the magnetic flux insertion in the ${\bf a}_2$-direction. (c) The finite-size scaling of the energy gap (green) and the ground-state splitting (red) for $N=4,5,6,8,10,12$ without flux insertions. We define the energy gap and the ground-state splitting as $E_4-E_1$ and $E_3-E_1$, respectively, where $E_i$ is the $i$th energy level in ascending order. (d) The particle entanglement spectrum (PES) for $N=10,N_1\times N_2=5\times 6,N_A=5$, with $23256$ levels below the entanglement gap (the dashed line).}
	\label{C_1_nu_1_3_bare}
\end{figure*}

\begin{figure*}[t]
	\centerline{\includegraphics[width=\linewidth]{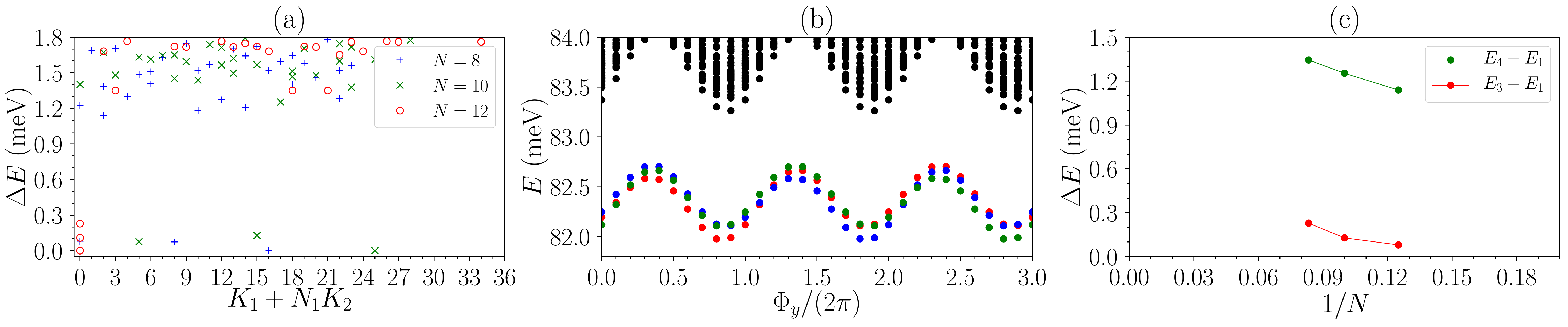}}
	\caption{Evidence of $\nu=1/3$ FCIs stabilized by the bare Coulomb interaction in the $\mathcal{C}=2$ region with $\theta=1.35^\circ$, $U=60{\rm meV}$. (a) The low-lying energy spectrum for $N=8, N_1\times N_2=4\times6$, $N=10,N_1\times N_2=5\times6$ and $N=12,N_1\times N_2=6\times6$. (b) The spectral flow for $N=10,N_1\times N_2=5\times 6$, where $\Phi_y$ is the magnetic flux insertion in the ${\bf a}_2$-direction. (c) The finite-size scaling of the energy gap (green) and the ground-state splitting (red) for $N=8,10,12$ without flux insertions. We define the energy gap and the ground-state splitting as $E_4-E_1$ and $E_3-E_1$, respectively, where $E_i$ is the $i$th energy level in ascending order. }
	\label{C_2_nu_1_3_bare}
\end{figure*}

\begin{figure*}[t]
\centerline{\includegraphics[width=\linewidth]{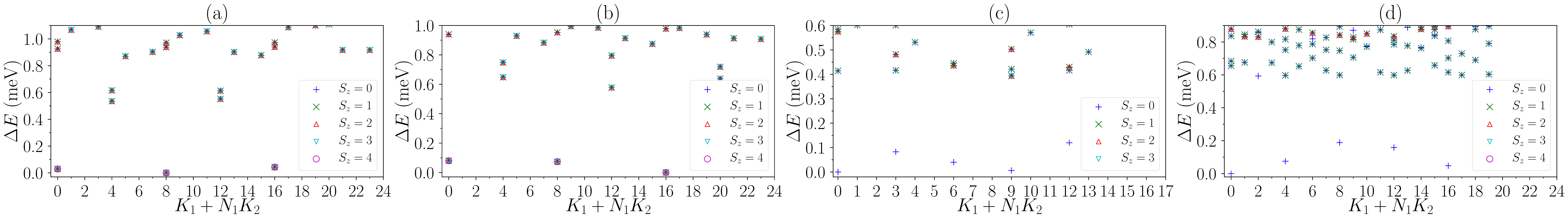}}
\caption{(a)-(b) The low-lying energy spectra of the bare Coulomb interaction at $\nu=1/3$ for $N=8$ spinful electrons on the $N_1\times N_2=4\times6$ lattice. The Chern number of the first conduction band is $\mathcal{C}=1$ in (a) and $\mathcal{C}=2$ in (b), with $\theta=0.75^\circ$, $U=20{\rm meV}$ and $\theta=1.35^\circ$, $U=60{\rm meV}$, respectively. (c)-(d) The low-lying energy spectra of the bare Coulomb interaction at $\nu=2/5$ for (c) $N=6$ spinful electrons on the $N_1\times N_2=3\times5$ lattice and (d) $N=8$ spinful electrons on the $N_1\times N_2=4\times5$ lattice with $\theta=0.65^\circ$, $U=20{\rm meV}$.}
\label{spin_bare}
\end{figure*}

In the main text, we use the Yukawa potential to describe the screening mechanism of the Coulomb interaction in TDBG. Strikingly, we obtain similar numerical results even by the unscreened bare Coulomb interaction. As shown in Figs.~\ref{C_1_nu_1_3_bare} and \ref{C_2_nu_1_3_bare}, under the assumption of both valley and spin polarization, robust $\nu=1/3$ FCIs can be stabilized by the bare Coulomb interaction in both the $\mathcal{C}=1$ and $\mathcal{C}=2$ regions. When the spin polarization is relaxed, the $\nu=1/3$ states are ferromagnetic in both $\mathcal{C}=1$ and $\mathcal{C}=2$ regions [Figs.~\ref{spin_bare}(a) and (b)]. Such ferromagnetism disappears at $\nu=2/5$, where we find the spin-singlet Halperin $(332)$ states are stabilized by the bare Coulomb interaction [Figs.~\ref{spin_bare}(c) and (d)].

\begin{figure*}[t!]
\centerline{\includegraphics[width=\linewidth]{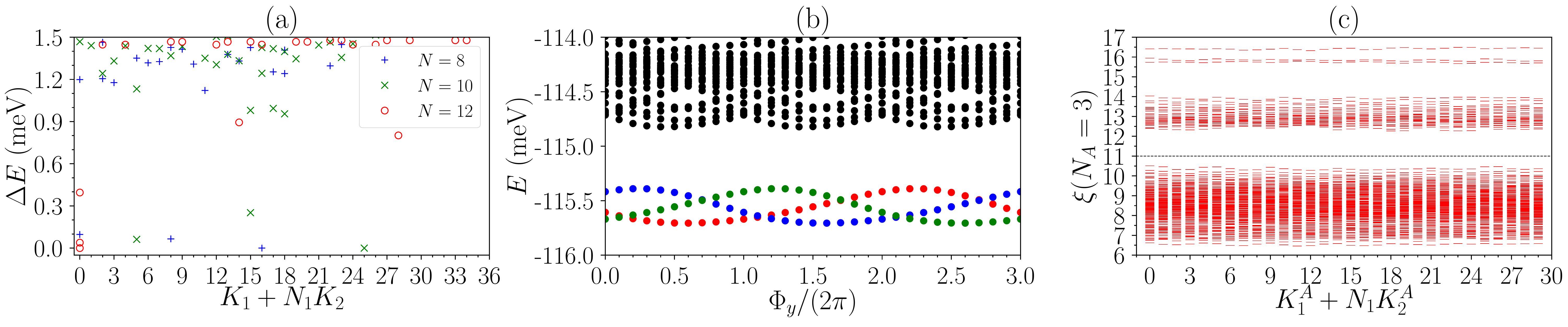}}
\caption{Evidence of $\nu=1/3$ FCIs stabilized by the bare Coulomb interaction in the $\mathcal{C}=2$ exactly flat band at the chiral limit. Here we choose $w_1=0.13{\rm eV}$. (a) The low-lying energy spectrum for $N=8, N_1\times N_2=4\times6$, $N=10,N_1\times N_2=5\times6$ and $N=12,N_1\times N_2=6\times6$. (b) The spectral flow for $N=10,N_1\times N_2=5\times 6$, where $\Phi_y$ is the magnetic flux insertion in the ${\bf a}_2$-direction. (c) The particle entanglement spectrum for $N=10,N_1\times N_2=5\times 6,N_A=3$, with $3250$ levels below the entanglement gap (the dashed line). The theoretical countings are $2530$ and $360$ for the $\nu=1/3$ Laughlin state and the charge density wave, respectively.}
\label{C_2_nu_1_3_chiral_bare}
\end{figure*}

In the language of Haldane's pseudopotential~\cite{HaldaneVm}, both the $\nu=1/3$ and $\nu=2/5$ FCIs are stabilized by short-range pseudopotentials. As the screening suppresses the long-range tail of the Coulomb interaction relative to the short-range part, we expect that these FCIs are not sensitive to the strength and details of the screening (although too strong screening suppresses the energy gap too much). Indeed, we obtain qualitatively the same numerical results by using another screening mechanism $V({\bf q})\propto\frac{1}{q}(1-e^{-q\kappa})$.

\subsection{Chiral limit}
Here we consider the chiral limit of TDBG by setting $t_3=t_4=\delta=U_1=U_2=w_0=0$ in Eq.~(\ref{smeq2}). At the magic angle $\theta=2\sin^{-1}\left(\frac{\sqrt{3}w_1}{4\pi\alpha t_0}\right)$ with $\alpha\approx 0.586$, there are two degenerate exactly flat bands at the charge neutrality. Such a degeneracy can be lifted by adding a diagonal term $\propto \mathds{1}\otimes\sigma_z$ in Eq.~(\ref{smeq1}), which gives $\mathcal{C}=\pm 2$ exactly flat bands below and above the charge neutrality. 

We now project the bare Coulomb interaction to this $\mathcal{C}=2$ exactly flat band. In this case, we also find nice three-fold ground-state degeneracies at $\nu=1/3$ [Figs.~\ref{C_2_nu_1_3_chiral_bare}(a)]. Such degeneracies can be further improved by increasing $w_1$. The spectral flow in Fig.~\ref{C_2_nu_1_3_chiral_bare} indicates that the ground-state manifold is robust against flux insertion. Moreover, we observe a clear entanglement gap in the PES [Fig.~\ref{C_2_nu_1_3_chiral_bare}(c)]. The number of levels below this gap exceeds the theoretical values of both the charge density waves and the $\nu=1/3$ Laughlin state, strongly suggesting a nontrivial topological order distinct from the $\nu=1/3$ Laughlin FCIs in $\mathcal{C}=1$ bands. We obtain very similar numerical results in the presence of screening of the Coulomb interaction.

As the $\mathcal{C}=2$ exactly flat band in the chiral limit is topologically equivalent to the $\mathcal{C}=2$ region of the realistic TDBG model considered in the main text, we expect the adiabatic continuity between the $\nu=1/3$ $\mathcal{C}=2$ FCIs shown in Fig.~\ref{C_2_nu_1_3_chiral_bare} and the $\nu=1/3$ $\mathcal{C}=2$ FCIs reported in the main text. Therefore, the $\nu=1/3$ $\mathcal{C}=2$ FCIs discovered in our realistic model should not be sensitive to the particular choice of the model parameters in Eqs.~(\ref{smeq1}) and (\ref{smeq2}), i.e., it can survive at smaller $t_3, t_4$ and $w_0$ which is closer to the chiral limit. In fact, it has been pointed out that smaller $t_3$ and $t_4$ is possible due to the lattice relaxation effect in TDBG~\cite{alexnanoletters}. Because the chiral limit has a variety of excellent analytical properties~\cite{alexnanoletters}, it could be an ideal platform to understand the nature of the $\nu=1/3$ $\mathcal{C}=2$ FCIs.

\end{document}